\documentclass[]{INCLUDES/llncs}
\usepackage{amsmath}
\usepackage{graphicx}
\usepackage{subfig}
\usepackage{url}
\usepackage{verbatim} 
\usepackage{listings}
\lstnewenvironment{codex}
    {\lstset{}%
      \csname lst@SetFirstLabel\endcsname}
    {\csname lst@SaveFirstLabel\endcsname}
\lstnewenvironment{code}
    {\lstset{}%
      \csname lst@SetFirstLabel\endcsname}
    {\csname lst@SaveFirstLabel\endcsname}
\newtheorem{prop}{Proposition}
\newtheorem{df}{Definition}

\newcommand{\BI}[0]{\begin{itemize}}
\newcommand{\EI}[0]{\end{itemize}}

\newcommand{\BE}[0]{\begin{enumerate}}
\newcommand{\EE}[0]{\end{enumerate}}

\newcommand{\BX}[0]{\begin{codex}}
\newcommand{\EX}[0]{\end{codex}}
\newcommand{\BV}[0]{\begin{verbatim}}
\newcommand{\EV}[0]{\end{verbatim}}

\def \bscale1 {1.00}
\def \bscale {0.20}

\begin{document}

\title{
  Pairing Functions, Boolean Evaluation and Binary Decision Diagrams
  in Prolog
}

\author{Paul Tarau}
\institute{
   Department of Computer Science and Engineering\\
   University of North Texas\\
   {\em E-mail: tarau@cs.unt.edu}
}

\maketitle

\date{}

\begin{abstract}

A ``pairing function'' J associates a unique natural number z to any two 
natural numbers x,y such that for two ``unpairing functions'' 
K and L, the equalities K(J(x,y))=x, L(J(x,y))=y and J(K(z),L(z))=z hold.
Using pairing functions on natural
number representations of truth tables, we derive 
an encoding for Binary Decision Diagrams with the unique property 
that its boolean evaluation faithfully mimics
its structural conversion to a a natural number
through recursive application of a matching pairing function.
We then use this result to derive {\em ranking} and {\em unranking}
functions for BDDs and reduced BDDs.
The paper is organized as a self-contained literate Prolog program,
available at \url{http://logic.csci.unt.edu/tarau/research/2008/pBDD.zip}.

\end{abstract}

\section{Introduction}

This paper is an exploration with logic programming tools of {\em ranking} and
{\em unranking} problems on Binary Decision Diagrams. The practical
expressiveness of logic programming languages (in particular Prolog) 
are put at test in the process. The paper is part
of a larger effort to cover in a declarative programming 
paradigm, arguably more elegantly, some fundamental combinatorial generation 
algorithms along the lines of \cite{knuth06draft}.
However, our main focus is by no means ``yet another implementation of BDDs in
Prolog''. The paper is more about fundamental
isomorphisms between logic functions
and their natural number representations, in the tradition of \cite{Goedel:31},
with the unusual twist that everything is expressed as a literate Prolog program,
and therefore automatically testable by the reader.
One could put such efforts under the generic umbrella of an emerging research
field that we would like to call {\em executable theoretical computer
science}. Nevertheless, we also hope that the more practically oriented reader 
will be able to benefit from this approach by being able to experiment with,
and reuse our Prolog code in applications.

The paper is organized as follows:
Sections \ref{bits} and \ref{bdds} overview efficient evaluation of boolean
formulae in Prolog using bitvectors represented as arbitrary length integers
and Binary Decision Diagrams (BDDs).
 
Section \ref{pairings} discusses classic pairing and unpairing
operations and introduces pairing/unpairing
predicates acting directly on bitlists.

Section \ref{encbdd} introduces a novel BDD encoding (based on our unpairing
functions) and discusses the surprising equivalence between boolean evaluation of BDDs
and the inverse of our encoding, the main result of the paper.

Section \ref{rank} describes {\em ranking} and {\em unranking}
functions for BDDs and reduced BDDs.

Sections \ref{related} and \ref{concl} discuss related work, 
future work and conclusions.

The code in the paper, embedded in a literate programming LaTeX
file, is entirely self contained and has been tested under {\em SWI-Prolog}.

\section{Parallel Evaluation of Boolean 
Functions with Bitvector Operations}\label{bits}

Evaluation of a boolean function can be performed one 
value at a time as in the predicate {\tt if\_then\_else/4}
\begin{code}
if_then_else(X,Y,Z,R):-
  bit(X),bit(Y),bit(Z),
  ( X==1->R=Y
  ; R=Z
  ).

bit(0).
bit(1).
\end{code}
\noindent resulting in a {\em truth table}\footnote{One can see that if the
number of variables is fixed, we can ignore the bitsrings in the brackets.
Thus, the truth table can be identified with the natural number, represented in
binary form by the last column.}
\begin{codex}
?- if_then_else(X,Y,Z,R),write([X,Y,Z]:R),nl,fail;nl.
[0, 0, 0]:0
[0, 0, 1]:1
[0, 1, 0]:0
[0, 1, 1]:1
[1, 0, 0]:0
[1, 0, 1]:0
[1, 1, 0]:1
[1, 1, 1]:1
\end{codex}
Clearly, this does not take advantage of the ability of modern hardware to
perform such operations one word a time - with the instant benefit of a
speed-up proportional to the word size.
An alternate representation, adapted
from \cite{knuth06draft} uses integer encodings 
of $2^n$ bits for each boolean variable $X_0,\ldots,X_{n-1}$. 
Bitvector operations evaluate all
value combinations at once.

\begin{prop}
Let $x_k$ be a variable for $0 \leq k<n$
where $n$ is the number of distinct variables in a 
boolean expression. Then column $k$ in the matrix representation
of the inputs in the the truth table
represents, as a bitstring, the natural number:

\begin{equation} \label{var}
x_k={(2^{2^n}-1)}/{(2^{2^{n-k-1}}+1)} 
\end{equation}
\end{prop}

\noindent For instance, if $n=2$, the formula computes 
$x_0=3=[0,0,1,1]$ and $x_1=5=[0,1,0,1]$.

The following predicates, working with arbitrary length bitstrings are
used to evaluate 
variables $x_k$ with $k \in [0..n-1]$ with formula \ref{var} 
and map the
constant boolean function {\tt 1} 
to the bitstring of length $2^n$, {\tt 111..1},
representing ${2^{2^n}}-1$
\begin{code}
% maps variable K in [0..NbOfBits-1] to Xk
var_to_bitstring_int(NbOfBits,K,Xk):-
  all_ones_mask(NbOfBits,Mask),
  var_to_bitstring_int(NbOfBits,Mask,K,Xk).

var_to_bitstring_int(NbOfBits,Mask,K,Xk):-
  NK is NbOfBits-(K+1),
  D is (1<<(1<<NK))+1,
  Xk is Mask//D.
  
% represents constant 1 as 11...1 build with NbOfBits bits 
all_ones_mask(NbOfBits,Mask):-Mask is (1<<(1<<NbOfBits))-1.   
\end{code}

We have used in {\tt var\_to\_bitstring\_int} an adaptation of the efficient 
bitstring-integer encoding described in the Boolean Evaluation 
section of \cite{knuth06draft}. Intuitively, it is based on the idea that one
can look at $n$ variables as bitstring representations of the $n$ columns
of the truth table.

Variables representing such bitstring-truth tables 
(seen as {\em projection functions}) 
can be combined with the usual bitwise integer operators, 
to obtain new bitstring truth tables, 
encoding all possible value combinations of their arguments.
Note that the constant $0$ is represented as $0$ while the constant $1$
is represented as $2^{2^n}-1$, corresponding to a column in
the truth table containing ones exclusively.
 
\section{Binary Decision Diagrams} \label {bdds}

We have seen that Natural Numbers in $[0..2^{2^n}-1]$ can be used as
representations of truth tables defining $n$-variable boolean functions.
A binary decision diagram (BDD)
\cite{bryant86graphbased} is an ordered binary tree obtained from 
a boolean function, by assigning its variables, one at a time, 
to {\tt 0}  (left branch) and {\tt 1} (right branch).
In virtually all practical applications BDDs are represented as DAGs after
detecting shared nodes. We safely ignore this here
as they represent the same logic
function, which is all we care about at this point. 
Typically in the early literature, the acronym
ROBDD is used to denote reduced ordered BDDs. Because this
optimization is now so prevalent, 
the term BDD is frequently use to refer to
ROBDDs. Strictly speaking, BDD in this paper will stand for {\em ordered BDD
with reduction of identical branches but without node sharing}.

The construction deriving a BDD of a boolean function $f$ is known as Shannon
expansion \cite{shannon_all}, and is expressed as

\begin{equation}
f(x)= (\bar{x} \wedge f[x \leftarrow 0]) \vee (x \wedge f[x \leftarrow 1])
\end{equation}

\noindent where $f[x \leftarrow a]$ is computed 
by uniformly substituting $a$ for $x$ in $f$. Note that by using the more
familiar boolean if-the-else function Shannon expansion can also
 be expressed as:

\begin{equation}
f(x) = if~x~then~f[x \leftarrow 1]~else~f[x \leftarrow 0]
\end{equation}

We represent a $BDD$ in Prolog as a binary tree with constants {\tt 0} and {\tt
1} as leaves, marked with the function symbol {\tt c/1}. Internal
{\em if-then-else} nodes marked with {\tt ite/3} are controlled by
variables, ordered identically in each branch, as first arguments of {\tt
ite/1}. The two other arguments are subtrees representing the {\tt Then} 
and {\tt Else} branches. Note that, in practice, reduced, 
canonical DAG representations are used instead of
binary tree representations.

Alternatively, we observe that the Shannon expansion
can be directly derived from a $2^n$ size truth table, 
using bitstring operations on encodings of its $n$ variables.
Assuming that the first column of a truth table corresponds to 
variable $x$, $x=0$ and $x=1$ mask out, respectively, 
the upper and lower half of the truth table.
 
\begin{code}
% splits a truth table of NV variables in 2 tables of NV-1 variables
shannon_split(NV,X, Hi,Lo):-
  all_ones_mask(NV,M),
  NV1 is NV-1,
  all_ones_mask(NV1,LM),
  HM is xor(M,LM),
  Lo is /\(LM,X),
  H is /\(HM,X),
  Hi is H>>(1<<NV1).
\end{code}
Note that the operation {\tt shannon\_split} can be reversed as follows:
\begin{code}
% fuses 2 truth tables of NV-1 variables into one of NV variables
shannon_fuse(NV,Hi,Lo, X):-
  NV1 is NV-1,
  H is Hi<<(1<<NV1),
  X is \/(H,Lo).
\end{code}
\begin{codex}
?- shannon_split(2, 7, X,Y),shannon_fuse(2, X,Y, Z).
X = 1,
Y = 3,
Z = 7.

?- shannon_split(3, 42, X,Y),shannon_fuse(3, X,Y, Z).
X = 2,
Y = 10,
Z = 42.
\end{codex}

Another way to look at these two operations (for a fixed value of NV), is
as bijections associating a pair of natural numbers to a 
natural number, i.e. as {\em pairing} functions.
\section{Pairing and Unpairing Functions} \label{pairings}

\begin{df}
A {\em pairing function} is a bijection $f : Nat \times Nat \rightarrow
Nat$. An {\em unpairing function} is a bijection $g : Nat \rightarrow
Nat  \times Nat$.
\end{df}

Following Julia Robinson's notation \cite{robinson50}, 
given a pairing function $J$, its left and right inverses $K$ and $L$ 
are such that

\begin{equation}
J(K(z),L(z))=z
\end{equation}

\begin{equation}
K(J(x,y))=x
\end{equation}

\begin{equation} 
L(J(x,y))=y 
\end{equation}

We refer to  \cite{DBLP:journals/tcs/CegielskiR01} for a typical use 
in the foundations of mathematics and to \cite{DBLP:conf/ipps/Rosenberg02a} 
for an extensive study of various pairing functions and their computational properties. 

\subsection{Cantor's Pairing Function}

Starting from Cantor's pairing function
\begin{code}
cantor_pair(K1,K2,P):-P is (((K1+K2)*(K1+K2+1))//2)+K2.
\end{code}
bijections from $Nat \times Nat$ to $Nat$ have been used for various proofs 
and constructions of mathematical objects 
\cite{robinson50,DBLP:journals/tcs/CegielskiR01}.

For $X,Y \in \{0,1,2,3\}$ the sequence of values of this pairing function is:
\begin{codex}
?- findall(R,(between(0,3,A),between(0,3,B),cantor_pair(A,B,R)),Rs).
Rs = [0, 2, 4, 6, 1, 5, 9, 13, 3, 11, 19, 27, 7, 23, 39, 55]
\end{codex}
\noindent Note however, that the inverse of Cantor's pairing function involves
potentially expensive floating point operations that are also likely to loose precision
for arbitrary length integers. 
\subsection{The Pepis-Kalmar Pairing Function}

Another pairing function that can be implemented using only
elementary integer operations is the following:

\begin{equation}
f(x,y)=2^x(2y+1)-1
\end{equation}

\noindent The predicates {\tt pepis\_pair/3} and {\tt pepis\_unpair/3} are
derived from the function {\bf pepis\_J} and its left and right unpairing 
companions {\bf pepis\_K} and {\bf pepis\_L} that have been used, by Pepis, 
Kalmar and Robinson 
in some fundamental work on recursion
theory, decidability and Hilbert's Tenth Problem 
in \cite{pepis,kalmar1,robinson67}:
\begin{code}
pepis_pair(X,Y,Z):-pepis_J(X,Y,Z).

pepis_unpair(Z,X,Y):-pepis_K(Z,X),pepis_L(Z,Y).
 
pepis_J(X,Y, Z):-Z is ((1<<X)*((Y<<1)+1))-1.
pepis_K(Z, X):-Z1 is Z+1,two_s(Z1,X).
pepis_L(Z, Y):-Z1 is Z+1,no_two_s(Z1,N),Y is (N-1)>>1. 

two_s(N,R):-even(N),!,H is N>>1,two_s(H,T),R is T+1.
two_s(_,0).

no_two_s(N,R):-two_s(N,T),R is N // (1<<T).

even(X):- 0 =:= /\(1,X).
odd(X):- 1 =:= /\(1,X).
\end{code}
This pairing function is asymmetrically growing
(faster growth on the first argument).
It works as follows:
\begin{codex}
?- pepis_pair(1,10,R).
R = 41.

?- pepis_unpair(10,1,R).
R = 3071.

?- findall(R,(between(0,3,A),between(0,3,B),pepis_pair(A,B,R)),Rs).
Rs=[0, 2, 4, 6, 1, 5, 9, 13, 3, 11, 19, 27, 7, 23, 39, 55]
\end{codex}

\subsection{Pairing/Unpairing 
operations acting directly on bitlists} \label{BitMerge}

We will describe here pairing operations, 
that are expressed exclusively as  bitlist transformations of
{\tt bitmerge\_unpair} and its inverse {\tt bitmerge\_pair},
and are therefore likely to be easily hardware implementable.
As we have found out recently, they turn out to be the same as the functions
defined in Steven Pigeon's PhD thesis on Data Compression \cite{pigeon}, page 114).

The predicate {\tt bitmerge\_pair} implements a bijection from $Nat \times
Nat$ to $Nat$ that works by splitting a number's big endian bitstring
representation into odd and even bits, while its inverse {\tt to\_pair} blends
the odd and even bits back together. The helper predicates 
{\tt to\_rbits} and {\tt from\_rbits}, 
given in the Appendix, convert to/from integers to bitlists.

\begin{code}
bitmerge_pair(X,Y,P):-
  to_rbits(X,Xs),
  to_rbits(Y,Ys),
  bitmix(Xs,Ys,Ps),!,
  from_rbits(Ps,P).

bitmerge_unpair(P,X,Y):-
  to_rbits(P,Ps),
  bitmix(Xs,Ys,Ps),!,
  from_rbits(Xs,X),
  from_rbits(Ys,Y).

bitmix([X|Xs],Ys,[X|Ms]):-!,bitmix(Ys,Xs,Ms).
bitmix([],[X|Xs],[0|Ms]):-!,bitmix([X|Xs],[],Ms).
bitmix([],[],[]).

\end{code}
The transformation of the bitlists, done by the bidirectional predicate bitmerge
is shown in the following example with bitstrings aligned:
\begin{codex}
?- bitmerge_unpair(2008,X,Y),bitmerge_pair(X,Y,Z).
X = 60,
Y = 26,
Z = 2008

% 2008:[0, 0, 0, 1, 1, 0, 1, 1, 1, 1, 1]
%   60:[      0,    1,    1,    1,    1]
%   26:[   0,    1,    0,    1,    1   ]
\end{codex}
Note that we represent numbers with bits in reverse order (least significant on
the left). Like in the case of Cantor's pairing function, we can see 
similar growth in both arguments:
\begin{codex}
?- between(0,15,N),bitmerge_unpair(N,A,B),
   write(N:(A,B)),write(' '),fail;nl.
0: (0, 0) 1: (1, 0) 2: (0, 1) 3: (1, 1) 
4: (2, 0) 5: (3, 0) 6: (2, 1) 7: (3, 1)
8: (0, 2) 9: (1, 2) 10: (0, 3) 11: (1, 3) 
12: (2, 2) 13: (3, 2) 14: (2, 3) 15: (3, 3)

?- between(0,3,A),between(0,3,B),bitmerge_pair(A,B,N),
   write(N:(A,B)),write(' '),fail;nl.
0: (0, 0) 2: (0, 1) 8: (0, 2) 10: (0, 3) 
1: (1, 0) 3: (1, 1) 9: (1, 2) 11: (1, 3) 
4: (2, 0) 6: (2, 1) 12: (2, 2) 14: (2, 3) 
5: (3, 0) 7: (3, 1) 13: (3, 2) 15: (3, 3)
\end{codex}
It is also convenient sometimes to see pairing/unpairing as one-to-one
functions from/to the underlying language's ordered pairs, i.e. {\tt X-Y} in
Prolog :
\begin{code}
bitmerge_pair(X-Y,Z):-bitmerge_pair(X,Y,Z).

bitmerge_unpair(Z,X-Y):-bitmerge_unpair(Z,X,Y).
\end{code}

\section{Encodings of Binary Decision Diagrams} \label{encbdd}

We will build a $BDD$ by applying {\tt bitmerge\_unpair}
recursively to a Natural Number {\tt TT}, 
seen as an $N$-variable $2^N$ bit truth table. 
This results in a complete binary tree of depth $N$.
As we will show later, this binary tree represents
a $BDD$ that returns {\tt TT} when evaluated applying
its boolean operations.

\begin{code}
% NV=number of varibles, TT=a truth table, BDD the result
plain_bdd(NV,TT, bdd(NV,BDD)):-
  Max is (1<<(1<<NV)),
  TT<Max,
  isplit(NV,TT, BDD).

% recurses to depth NV, splitting TT into pairs
isplit(0,TT,c(TT)).
isplit(NV,TT,R):-NV>0,
  NV1 is NV-1,
  bitmerge_unpair(TT,Hi,Lo),
  isplit(NV1,Hi,H),
  isplit(NV1,Lo,L),
  ite(NV1,H,L)=R.
\end{code}
The following examples 
show the results returned by {\tt plain\_bdd} 
for all $2^{2^k}$ truth tables associated to $k$ variables,  with $k=2$.

\begin{codex}
?- between(0,15,TT),plain_bdd(2,TT,BDD),write(TT:BDD),nl,fail;nl
0:bdd(2, ite(1, ite(0, c(0), c(0)), ite(0, c(0), c(0))))
1:bdd(2, ite(1, ite(0, c(1), c(0)), ite(0, c(0), c(0))))
2:bdd(2, ite(1, ite(0, c(0), c(0)), ite(0, c(1), c(0))))
...
13:bdd(2, ite(1, ite(0, c(1), c(1)), ite(0, c(0), c(1))))
14:bdd(2, ite(1, ite(0, c(0), c(1)), ite(0, c(1), c(1))))
15:bdd(2, ite(1, ite(0, c(1), c(1)), ite(0, c(1), c(1))))
\end{codex}

\subsection{Reducing the $BDDs$}
The predicate {\tt bdd\_reduce} reduces a $BDD$ by trimming identical 
left and right subtrees, and the predicate {\tt bdd} 
associates this reduced form to $N \in Nat$.
\begin{code}
bdd_reduce(BDD,bdd(NV,R)):-nonvar(BDD),BDD=bdd(NV,X),bdd_reduce1(X,R).

bdd_reduce1(c(TT),c(TT)).
bdd_reduce1(ite(_,A,B),R):-A==B,bdd_reduce1(A,R).
bdd_reduce1(ite(X,A,B),ite(X,RA,RB)):-A\==B,
  bdd_reduce1(A,RA),bdd_reduce1(B,RB).

bdd(NV,TT, ReducedBDD):-
  plain_bdd(NV,TT, BDD),
  bdd_reduce(BDD,ReducedBDD).
\end{code}
Note that we omit here the reduction step consisting in
sharing common subtrees, as it is obtained easily by replacing
trees with DAGs. The process is facilitated by the fact
that our unique encoding provides a perfect hashing
key for each subtree. The following examples 
show the results returned by {\tt bdd} for {\tt NV=2}.

\begin{codex}
?- between(0,15,TT),bdd(2,TT,BDD),write(TT:BDD),nl,fail;nl
0:bdd(2, c(0))
1:bdd(2, ite(1, ite(0, c(1), c(0)), c(0)))
2:bdd(2, ite(1, c(0), ite(0, c(1), c(0))))
3:bdd(2, ite(0, c(1), c(0)))
...
13:bdd(2, ite(1, c(1), ite(0, c(0), c(1))))
14:bdd(2, ite(1, ite(0, c(0), c(1)), c(1)))
15:bdd(2, c(1))
\end{codex}

\subsection{From BDDs to Natural Numbers}
One can ``evaluate back'' the binary tree representing the BDD,
by using the pairing function {\tt bitmerge\_pair}.  
The inverse of {\tt plain\_bdd} is implemented as follows:
\begin{code}
plain_inverse_bdd(bdd(_,X),TT):-plain_inverse_bdd1(X,TT).

plain_inverse_bdd1(c(TT),TT).
plain_inverse_bdd1(ite(_,L,R),TT):-
  plain_inverse_bdd1(L,X),
  plain_inverse_bdd1(R,Y),
  bitmerge_pair(X,Y,TT).
\end{code}

\begin{codex}
?- plain_bdd(3,42, BDD),plain_inverse_bdd(BDD,N).
BDD = bdd(3, 
          ite(2, 
              ite(1, 
                  ite(0, c(0), c(0)), 
                  ite(0, c(0), c(0))), 
              ite(1, 
                  ite(0, c(1), c(1)), 
                  ite(0, c(1), c(0))))),
N = 42
\end{codex}
\noindent Note however that {\tt plain\_inverse\_bdd/2} does not act as an
inverse of {\tt bdd/3}, given that the {\em structure} of the $BDD$ tree 
is changed by reduction.

\subsection{Boolean Evaluation of BDDs}
This raises the obvious question: how can we recover the original truth
table from a reduced BDD? The obvious answer is: by evaluating it as a
boolean function! The predicate {\tt ev/2} describes the $BDD$ evaluator:
\begin{code}
ev(bdd(NV,B),TT):-
  all_ones_mask(NV,M),
  eval_with_mask(NV,M,B,TT).

evc(0,_,0).
evc(1,M,M).

eval_with_mask(_,M,c(X),R):-evc(X,M,R).
eval_with_mask(NV,M,ite(X,T,E),R):-
  eval_with_mask(NV,M,T,A),
  eval_with_mask(NV,M,E,B),
  var_to_bitstring_int(NV,M,X,V),
  ite(V,A,B,R).
\end{code}
The predicate {\tt ite/4} used in {\tt eval\_with\_mask} 
implements the boolean function  {\tt if X then T else E}
using arbitrary length bitvector operations:
\begin{code}
ite(X,T,E, R):-R is xor(/\(X,xor(T,E)),E).
\end{code}
Note that this equivalent formula for {\tt ite} is slightly more
efficient than the obvious one with $\wedge$ and $\vee$ as it
requires only $3$ boolean operations. We will use {\tt ite/4} as the
basic building block for implementing a boolean evaluator for BDDs.

\subsection{The Equivalence}
A surprising result
is that boolean evaluation and structural transformation with
repeated application of
{\em pairing}
produce the same result, i.e. 
the predicate {\tt ev/2} also acts as an inverse 
of {\tt bdd/2} and {\tt plain\_bdd/2}.

\noindent {\em 
As the following example shows, boolean evaluation {\tt ev/2}
faithfully emulates {\tt plain\_inverse\_bdd/2}, 
on both plain and reduced BDDs.
}

\begin{codex}
?- plain_bdd(3,42,BDD),ev(BDD,N).
BDD = bdd(3, 
        ite(2, 
            ite(1, 
                ite(0, c(0), c(0)), 
                ite(0, c(0), c(0))), 
            ite(1, 
                ite(0, c(1), c(1)), 
                ite(0, c(1), c(0))))),
N = 42

?- bdd(3,42,BDD),ev(BDD,N).
BDD = bdd(3, 
         ite(2, 
            c(0), 
            ite(1, 
                c(1), 
                ite(0, c(1), c(0))))),
N = 42
\end{codex}

The main result of this subsection can now be summarized as follows:
\begin{prop} \label{tt}
Let $B$ be the complete binary tree of depth $N$, obtained by recursive 
applications of {\tt bitmerge\_unpair} on a truth table $T$, as described
by the predicate {\tt plain\_bdd(N,T,B)}.

Then for any $N$ and any $T$, when $B$ is interpreted as an (unreduced) BDD,
the result $V$ of its boolean evaluation using the predicate $ev(N,B,V)$ 
and
the result $R$ obtained by applying $plain\_inverse\_bdd(N,B,R)$ 
are both identical to $T$. Moreover, the operation {\tt $ev(N,B,V)$}
reverses the effects of both {\tt plain\_bdd} and {\tt bdd} with an 
identical result.
\end{prop}

\noindent {\em Proof:} The predicate {\tt plain\_bdd} builds a binary 
tree by splitting the bitstring $tt \in [0..2^N-1]$ up to depth $N$. 
Observe that this corresponds to the Shannon expansion \cite{shannon_all} of the
formula associated to the truth table, using variable order $[n-1,...,0]$.
Observe that the effect of {\tt bitstring\_unpair} is the same as
\begin{itemize}
  \item the effect of {\tt var\_to\_bitstring\_int(N,M,(N-1),R)} 
     acting as a mask selecting the left branch
\item 
     and the effect of its complement, acting as a mask selecting the right
     branch.
\end{itemize}
Given that $2^N$ is the double of $2^{N-1}$, the same invariant holds at each step, 
as the bitstring length of the truth table reduces to half. On the other hand,
it is clear that {\tt $ev$} reverses the action of both {\tt plain\_bdd} and
{\tt bdd} as BDDs and reduced BDDs represent 
the same boolean function \cite{bryant86graphbased}.

This result can be seen as a yet another intriguing isomorphism between
boolean, arithmetic and symbolic computations.

\section{Ranking and Unranking of BDDs} \label{rank}


One more step is needed to extend the mapping between $BDDs$ with $N$
variables to a bijective mapping from/to $Nat$: 
we will have to ``shift toward infinity'' 
the starting point of each new block of 
BDDs in $Nat$ as BDDs of larger and larger sizes are enumerated.

First, we need to know by how much - so we compute the sum of the
counts of boolean functions with up to $N$ variables.

\begin{code}
bsum(0,0).
bsum(N,S):-N>0,N1 is N-1,bsum1(N1,S).

bsum1(0,2).
bsum1(N,S):-N>0,N1 is N-1,bsum1(N1,S1),S is S1+(1<<(1<<N)).
\end{code}

The stream of all such sums can now be generated as usual:
\begin{code}
bsum(S):-nat(N),bsum(N,S).

nat(0).
nat(N):-nat(N1),N is N1+1.
\end{code}
What we are really interested in, is decomposing {\tt N} into
the distance to the
last {\tt bsum} smaller than N, {\tt N\_M}
and the index of that generates the sum, {\tt K}.
\begin{code}
to_bsum(N, X,N_M):-
  nat(X),bsum(X,S),S>N,!,
  K is X-1,
  bsum(K,M),
  N_M is N-M.
\end{code}
{\em Unranking} of an arbitrary BDD is now easy - the index {\tt K}
determines the number of variables and {\tt N\_M} determines
the rank. Together they select the right BDD
with {\tt plain\_bdd} and {\tt bdd/3}.
\begin{code}
nat2plain_bdd(N,BDD):-to_bsum(N, K,N_M),plain_bdd(K,N_M,BDD).

nat2bdd(N,BDD):-to_bsum(N, K,N_M),bdd(K,N_M,BDD).
\end{code}
{\em Ranking} of a BDD is even easier: we first compute
its {\tt NumberOfVars} and its rank {\tt Nth}, then we shift the rank by 
the {\tt bsums} up to {\tt NumberOfVars}, enumerating the
ranks previously assigned.
\begin{code}
plain_bdd2nat(bdd(NumberOfVars,BDD),N) :-
  B=bdd(NumberOfVars,BDD),
  plain_inverse_bdd(B,Nth),
  K is NumberOfVars-1,
  bsum(K,S),N is S+Nth.

bdd2nat(bdd(NumberOfVars,BDD),N) :-
  B=bdd(NumberOfVars,BDD),
  ev(B,Nth),
  K is NumberOfVars-1,
  bsum(K,S),N is S+Nth.  
\end{code}
As the following example shows, {\tt nat2plain\_bdd/2}
and {\tt plain\_bdd2nat/2} implement inverse functions.
\begin{codex}
?- nat2plain_bdd(42,BDD),plain_bdd2nat(BDD,N).
BDD = bdd(4, 
          ite(3, 
              ite(2, 
                  ite(1, 
                      ite(0, c(0), c(0)), 
                      ite(0, c(1), c(0))), 
                  ite(1, 
                      ite(0, c(1), c(0)), 
                      ite(0, c(0), c(0)))), 
              ite(2, 
                  ite(1, 
                      ite(0, c(0), c(0)), 
                      ite(0, c(0), c(0))), 
                  ite(1, ite(0, c(0), c(0)), 
                         ite(0, c(0), c(0)))))),
N = 42
\end{codex}
\noindent The same applies to {\tt nat2bdd/2} and its 
inverse {\tt bdd2nat/2}.
\begin{codex}
?- nat2bdd(42,BDD),bdd2nat(BDD,N).
BDD = bdd(4, 
          ite(3, 
              ite(2, 
              ite(1, c(0), 
                     ite(0, c(1), c(0))), 
                     ite(1, 
                         ite(0, c(1),c(0)), c(0))), 
                         c(0))),
N = 42
\end{codex}
\noindent We can now generate infinite streams of BDDs as follows:
\begin{code}
plain_bdd(BDD):-nat(N),nat2plain_bdd(N,BDD).

bdd(BDD):-nat(N),nat2bdd(N,BDD).
\end{code}
\begin{codex}
?- plain_bdd(BDD).
BDD = bdd(1, ite(0, c(0), c(0))) ;
BDD = bdd(1, ite(0, c(1), c(0))) ;
BDD = bdd(2, ite(1, ite(0, c(0), c(0)), ite(0, c(0), c(0)))) ;
BDD = bdd(2, ite(1, ite(0, c(1), c(0)), ite(0, c(0), c(0)))) ;
...
?- bdd(BDD).
BDD = bdd(1, c(0)) ;
BDD = bdd(1, ite(0, c(1), c(0))) ;
BDD = bdd(2, c(0)) ;
BDD = bdd(2, ite(1, ite(0, c(1), c(0)), c(0))) ;
BDD = bdd(2, ite(1, c(0), ite(0, c(1), c(0)))) ;
BDD = bdd(2, ite(0, c(1), c(0))) ;
...
\end{codex}

\section{Related work} \label{related}
Pairing functions have been used in work on decision problems as early 
as \cite{pepis,kalmar1,robinson50}.
{\em Ranking} functions can be traced back to G\"{o}del numberings
\cite{Goedel:31,conf/icalp/HartmanisB74} associated to formulae. 
Together with their inverse {\em unranking} functions they are also 
used in combinatorial generation
algorithms \cite{conf/mfcs/MartinezM03,knuth06draft}. 
Binary Decision Diagrams are the dominant boolean function representation in
the field of circuit design automation
\cite{DBLP:journals/tcad/DrechslerSF04}.
BDDs have been used in a Genetic Programming context
\cite{DBLP:conf/ices/SakanashiHIK96,DBLP:journals/heuristics/ChenLHW04}
as a representation of evolving individuals subject to crossovers and mutations expressed as
structural transformations and recently in a machine learning context for
compressing probabilistic Prolog programs \cite{DBLP:journals/ml/RaedtKKRT08}
representing candidate theories. 
Other interesting uses of BDDs in a 
logic and constraint programming context are 
related to representations of
finite domains. In \cite{DBLP:conf/padl/HawkinsS06} an algorithm for
finding minimal reasons for inferences is given.

\section{Conclusion and Future Work} \label{concl}
The surprising connection of pairing/unpairing functions and BDDs, 
is the indirect result of implementation
work on a number of practical applications.
Our initial interest has been triggered by applications of the 
encodings to combinational circuit synthesis in a logic 
programming framework \cite{cf08,iclp07}.
We have found them also interesting as uniform 
blocks for Genetic Programming applications of Logic Programming.
In a Genetic Programming context \cite{koza92}, 
the bijections between bitvectors/natural numbers 
on one side, and trees/graphs representing BDDs on the other side, 
suggest exploring the mapping and its action on various
transformations as a phenotype-genotype connection. 
Given the connection between BDDs to
boolean and finite domain constraint solvers
it would be interesting to explore in that context,
efficient succinct data representations
derived from our BDD encodings.

\bibliographystyle{INCLUDES/splncs}
\bibliography{INCLUDES/theory,tarau,INCLUDES/proglang,INCLUDES/biblio,INCLUDES/syn}

\begin{thebibliography}{10}

\bibitem{knuth06draft}
Knuth, D.:
\newblock {The Art of Computer Programming, Volume 4, draft} (2006)
  http://www-cs-faculty.stanford.edu/$\sim$knuth/taocp.html.

\bibitem{Goedel:31}
G\"{o}del, K.:
\newblock \"{U}ber formal unentscheidbare {S\"{a}tze der Principia Mathematica
  und verwandter Systeme I}.
\newblock Monatshefte f\"{u}r Mathematik und Physik \textbf{38} (1931)
  173--198

\bibitem{bryant86graphbased}
Bryant, R.E.:
\newblock Graph-based algorithms for boolean function manipulation.
\newblock {IEEE} Transactions on Computers \textbf{35}(8) (1986)  677--691

\bibitem{shannon_all}
Shannon, C.E.:
\newblock {Claude Elwood Shannon: collected papers}.
\newblock IEEE Press, Piscataway, NJ, USA (1993)

\bibitem{robinson50}
Robinson, J.:
\newblock General recursive functions.
\newblock Proceedings of the American Mathematical Society \textbf{1}(6) (dec
  1950)  703--718

\bibitem{DBLP:journals/tcs/CegielskiR01}
C{\'e}gielski, P., Richard, D.:
\newblock Decidability of the theory of the natural integers with the cantor
  pairing function and the successor.
\newblock Theor. Comput. Sci. \textbf{257}(1-2) (2001)  51--77

\bibitem{DBLP:conf/ipps/Rosenberg02a}
Rosenberg, A.L.:
\newblock Efficient pairing functions - and why you should care.
\newblock International Journal of Foundations of Computer Science
  \textbf{14}(1) (2003)  3--17

\bibitem{pepis}
Pepis, J.:
\newblock Ein verfahren der mathematischen logik.
\newblock The Journal of Symbolic Logic \textbf{3}(2) (jun 1938)  61--76

\bibitem{kalmar1}
Kalmar, L.:
\newblock On the reduction of the decision problem. first paper. ackermann
  prefix, a single binary predicate.
\newblock The Journal of Symbolic Logic \textbf{4}(1) (mar 1939)  1--9

\bibitem{robinson67}
Robinson, J.:
\newblock An introduction to hyperarithmetical functions.
\newblock The Journal of Symbolic Logic \textbf{32}(3) (sep 1967)  325--342

\bibitem{pigeon}
Pigeon, S.:
\newblock Contributions \`{a} la compression de donn\'{e}es.
\newblock Ph.d. thesis, Universit\'{e} de Montr\'{e}al, Montr\'{e}al (2001)

\bibitem{conf/icalp/HartmanisB74}
Hartmanis, J., Baker, T.P.:
\newblock On simple goedel numberings and translations.
\newblock In Loeckx, J., ed.: ICALP. Volume~14 of Lecture Notes in Computer
  Science., Springer (1974)  301--316

\bibitem{conf/mfcs/MartinezM03}
Martinez, C., Molinero, X.:
\newblock Generic algorithms for the generation of combinatorial objects.
\newblock In Rovan, B., Vojtas, P., eds.: MFCS. Volume 2747 of Lecture Notes in
  Computer Science., Springer (2003)  572--581

\bibitem{DBLP:journals/tcad/DrechslerSF04}
Drechsler, R., Shi, J., Fey, G.:
\newblock Synthesis of fully testable circuits from bdds.
\newblock IEEE Trans. on CAD of Integrated Circuits and Systems \textbf{23}(3)
  (2004)  440--443

\bibitem{DBLP:conf/ices/SakanashiHIK96}
Sakanashi, H., Higuchi, T., Iba, H., Kakazu, Y.:
\newblock Evolution of binary decision diagrams for digital circuit design
  using genetic programming.
\newblock In Higuchi, T., Iwata, M., Liu, W., eds.: ICES. Volume 1259 of
  Lecture Notes in Computer Science., Springer (1996)  470--481

\bibitem{DBLP:journals/heuristics/ChenLHW04}
Chen, S.T., Lin, S.S., Huang, L.T., Wei, C.J.:
\newblock Towards the exact minimization of bdds-an elitism-based distributed
  evolutionary algorithm.
\newblock J. Heuristics \textbf{10}(3) (2004)  337--355

\bibitem{DBLP:journals/ml/RaedtKKRT08}
Raedt, L.D., Kersting, K., Kimmig, A., Revoredo, K., Toivonen, H.:
\newblock Compressing probabilistic prolog programs.
\newblock Machine Learning \textbf{70}(2-3) (2008)  151--168

\bibitem{DBLP:conf/padl/HawkinsS06}
Hawkins, P., Stuckey, P.J.:
\newblock A hybrid bdd and sat finite domain constraint solver.
\newblock In Hentenryck, P.V., ed.: PADL. Volume 3819 of Lecture Notes in
  Computer Science., Springer (2006)  103--117

\bibitem{cf08}
Tarau, P., Luderman, B.:
\newblock Exact combinational logic synthesis and non-standard circuit design.
\newblock In: CF '08: Proceedings of the 2008 conference on Computing
  frontiers, New York, NY, USA, ACM (2008)  179--188

\bibitem{iclp07}
Tarau, P., Luderman, B.:
\newblock {A Logic Programming Framework for Combinational Circuit Synthesis}.
\newblock In: {23rd International Conference on Logic Programming (ICLP), LNCS
  4670}, Porto, Portugal, Springer (September 2007)  180--194

\bibitem{koza92}
Koza, J.R.:
\newblock Genetic Programming: On the Programming of Computers by Means of
  Natural Selection.
\newblock MIT Press, Cambridge, MA, USA (1992)

\end{thebibliography}

\subsection*{Appendix}
To make the code in the paper fully self contained, 
we list here some auxiliary functions.

\paragraph{Bit crunching operations}
\begin{code}
% converts an int to a list of bits, least significant first
to_rbits(0,[]).
to_rbits(N,[B|Bs]):-N>0,B is N mod 2, N1 is N//2,
  to_rbits(N1,Bs).

% converts a list of bits (least significant first) into an int
from_rbits(Rs,N):-nonvar(Rs),from_rbits(Rs,0,0,N).

from_rbits([],_,N,N).
from_rbits([X|Xs],E,N1,N3):-NewE is E+1,N2 is X<<E+N1,
  from_rbits(Xs,NewE,N2,N3).
\end{code}
\end{document}